\documentclass[twocolumn]{aastex631}

\received{July 21, 2021}
\revised{October 8, 2021}
\accepted{October 25, 2021}
\submitjournal{ApJ}

\shorttitle{QUBRICS Ultra-luminous QSOs at $z\sim 5$}
\shortauthors{Grazian et al.}
\graphicspath{{./}{figures/}}

\begin{document}

\title{The space density of ultra-luminous QSOs at the end of
reionization epoch by the QUBRICS Survey and the AGN contribution to
the hydrogen ionizing background}

\correspondingauthor{Andrea Grazian}
\email{andrea.grazian@inaf.it}

\author{Andrea Grazian}
\affil{INAF--Osservatorio Astronomico di Padova, 
Vicolo dell'Osservatorio 5, I-35122, Padova, Italy\\}

\author{Emanuele Giallongo}
\affil{INAF--Osservatorio Astronomico di Roma, Via Frascati 33, I-00078,
Monte Porzio Catone, Italy}

\author{Konstantina Boutsia}
\affil{Las Campanas Observatory, Carnegie Observatories, 
Colina El Pino, Casilla 601, La Serena, Chile\\}

\author{Giorgio Calderone}
\affil{INAF--Osservatorio Astronomico di Trieste, 
Via G.B. Tiepolo, 11, I-34143, Trieste, Italy \\}

\author{Stefano Cristiani}
\affil{INAF--Osservatorio Astronomico di Trieste, 
Via G.B. Tiepolo, 11, I-34143, Trieste, Italy \\}
\affiliation{INFN-National Institute for Nuclear Physics,  
via Valerio 2, I-34127 Trieste \\}
\affil{IFPU--Institute for Fundamental Physics of the Universe,
via Beirut 2, I-34151, Trieste, Italy}

\author{Guido Cupani}
\affil{INAF--Osservatorio Astronomico di Trieste, 
Via G.B. Tiepolo, 11, I-34143, Trieste, Italy \\}
\affil{IFPU--Institute for Fundamental Physics of the Universe,
via Beirut 2, I-34151, Trieste, Italy}

\author{Fabio Fontanot}
\affil{INAF--Osservatorio Astronomico di Trieste, 
Via G.B. Tiepolo, 11, I-34143, Trieste, Italy \\} 
\affil{IFPU--Institute for Fundamental Physics of the Universe,
via Beirut 2, I-34151, Trieste, Italy}

\author{Francesco Guarneri}
\affil{INAF--Osservatorio Astronomico di Trieste, 
Via G.B. Tiepolo, 11, I-34143, Trieste, Italy \\}
\affil{Dipartimento di Fisica, Sezione di Astronomia,
Universit\`a di Trieste, via G.B. Tiepolo 11, I-34131, Trieste, Italy}

\author{Yacob Ozdalkiran}
\affil{Ecole Polytechnique Paris, Rte de Saclay, F-91120, Palaiseau, France}

\begin{abstract}
Motivated by evidences favoring a rapid and late hydrogen
reionization process completing at $z\sim 5.2-5.5$ and mainly driven
by rare and luminous sources, we have reassessed the estimate of the
space density of ultra-luminous QSOs at $z\sim 5$ in the framework of
the QUBRICS survey. A $\sim 90\%$ complete sample of 14
spectroscopically confirmed QSOs at $M_{1450}\le -28.3$ and $4.5\le
z\le 5.0$ has been derived in an area of $12,400$ sq. deg., thanks to
multi-wavelength selection and GAIA astrometry. The space density of $z\sim 5$ QSOs
within $-29.3\le M_{1450}\le -28.3$ is three times higher than
previous determinations. Our
results suggest a steep bright-end slope for the QSO luminosity
function at $z\sim 5$ and a mild redshift evolution of the space
density of ultra-bright QSOs ($M_{1450}\sim -28.5$) at $3<z<5.5$, in
agreement with the redshift evolution of the much fainter AGN
population at $M_{1450}\sim -23$. These findings are consistent with a
pure density evolution for the AGN population at $z>3$. Adopting our
$z\sim 4$ QSO luminosity function and applying a mild density
evolution in redshift, a photo-ionization rate of
$\Gamma_{HI}=0.46^{+0.17}_{-0.09}\times 10^{-12}s^{-1}$ has been
obtained at $z=4.75$, assuming an escape fraction of $\sim 70\%$
and a steep faint-end slope of the AGN luminosity function.
The derived photo-ionization rate is $\sim 50-100\%$ of the ionizing
background measured at the end of the reionization epoch, suggesting
that AGNs could play an important role in the cosmological reionization process.
\end{abstract}


\section{Introduction} \label{sec:intro}

At $z<10$, the Universe underwent a ``disruptive'' phase transition,
usually called Reionization, causing the neutral hydrogen fraction
to drop from unity to a value of $\sim 10^{-4}$ at $z<5-6$, after
which the Universe was almost completely ionized \citep{fan06}. The low optical
depth of the cosmic microwave background (CMB) photons measured by the
Planck satellite \citep{planck20} supports such a scenario with a
midpoint redshift $z\lesssim 7$.
However, the timing and topology of reionization is still unclear,
especially because we do not know the relative contribution of
star-forming galaxies and AGNs to the hydrogen (HI) ionizing background
in the Universe close to the reionization epoch.
Moreover, while in the past few years this process was thought to start
early and developing in a relatively large redshift interval $6<z<15$
\citep[e.g.][]{bouwens09}, there are now accumulating undisputed evidences
in favor of a late and short reionization process.

The strong redshift evolution of the mean free path (MFP) of ionizing
photons into the intergalactic medium (IGM), which appears to decrease
significantly in the redshift interval $4<z<6$, could be indicative of
a rapid change in the ionization level of the IGM since, for a given
source emissivity, the photo-ionization rate $\Gamma_{HI}$ is simply
proportional to the MFP. Indeed a fast drop of the MFP at
$z\sim 6$ has been recently derived by \citet{becker21} with respect
to the extrapolation from values measured at lower redshifts \citep{worseck14}.

Additional support to a quick and late reionization scenario is provided by
several independent observations, including (1) the rapid drop of the space
density of Lyman-$\alpha$ emitters (LAEs) at $z>6$ \citep{morales21},
(2) the rapid
decrease of the fraction of LAEs within the Lyman-break galaxy
population at $z\sim 6-7$ \citep{hoag19}, (3) the detection of
the patchy kinetic Sunyaev-Zeldovich (kSZ) effect, indicating a short
duration phase $\Delta z\le 1-2.8$ of the reionization event
\citep{george15,reichardt20}. All these observational evidences suggest
that the $z\gtrsim 7$ IGM should be almost neutral, with $x_{HI}>0.6$.
This implies a rapid reionization process which is effective in the
short redshift interval $5\lesssim z\lesssim 6.5$.
In this rapid and late reionization scenario, sparse over-densities of
neutral hydrogen are still present at redshifts as low as $z\sim 5.2$
in the diffuse ionized IGM, implying an end of reionization at
$z\sim 5$ \citep{keating20,bosman21,zhu21}.

A rapid and late reionization scenario is in tension with models assuming
galaxies as the only contributors to the budget of HI ionizing background.
Indeed, models assuming
star forming galaxies as main contributors to the UV background show
that (relatively faint) galaxies tend to start the reionization
process too early \citep{naidu20}. It is also worth noting that UV
selected star forming galaxies do not show an abrupt drop in their UV
luminosity density at $z\sim 5.5$, where the bulk of the reionization
event is taking place, but they show an accelerated evolution only at
$z\sim 8-9$ \citep{oesch18,bouwens21}, too early to be in agreement
with the redshift evolution shown by the MFP measured
by \citet{becker21} and by the photo-ionization rate
\citep{calverley11,wyithe11,davies18}.

Moreover, an escape fraction of ionizing photons from galaxy halos of
$f_{esc}\sim 10$\% for the global galaxy population down to $M_{1500}\sim -12$
\citep{kuhlen12,robertson15,finkelstein19} would
be needed to accomplish the rapid evolution of the MFP.
Such a value for $f_{esc}$ is much
higher than what is directly measured in galaxies at lower redshifts
\citep[$z=3-4$; e.g.][]{grazian16,grazian17,pahl20}. This
tension would require rapid evolution in the physical properties of
the global galaxy population \citep{davies21} or relaxing somewhat the
constraints derived from the MFP measurements \citep{cain21}. The
difficulty in ascribing the photon budget, needed to drive this quick
and late reionization, to the global galaxy population alone, leaves
room to a significant role of the AGN population in providing a
significant photo-ionization rate at late cosmic times ($z<6.5$) and an
acceleration to the reionization process.

AGNs indeed are known to ionize their environment, producing a
remarkable proximity effect in their Lyman-$\alpha$ forest absorption
spectra. This enhancement in the ionization level of the IGM due to the
ionizing QSO flux extends to several proper Mpc ($\sim 6-10$ pMpc) as
observed in bright QSOs
\citep[$M_{1450}\lesssim -27$;][]{fan06,calverley11,eilers17}.
This implies that their
escape fraction is in general relatively large ($\sim 70$\%) and
appears to keep values $>50$\% in fainter AGNs
\citep[$M_{1450}\lesssim -23$;][]{cristiani16,grazian18,romano19}.
Thus, the relevance of AGNs
in the reionization process mainly depends on the space density
evolution of AGNs with $M_{1450}< -22$ in the redshift interval
$3.5<z<6.5$.
However, the assessment of their space density at very high redshifts
is challenging both for relatively bright QSOs and fainter AGNs,
requiring a wide multi-wavelength approach and ancillary selection
techniques derived e.g. by the astrometric and/or variability analyses.

At $z\sim 4-4.5$ there is an agreement among different spectroscopic
and photometric surveys based on multi-wavelength data, suggesting a
double power-law shape for the luminosity function with a break
at $M_{1450}\sim -26\div -26.5$
\citep{glikman11,boutsia18,giallongo19,boutsia21} and a steep
bright-end slope $\beta \sim -3.7\div -4$.
This luminosity function is higher with respect to
previous standard color-selected surveys \citep{akiyama18,Sch19a,Sch19b},
both at bright and intermediate magnitudes near the break.

While at redshift $z\gtrsim 6$ we have only scanty data mainly
confined to bright QSOs, an increasing data-set coming from
spectroscopic and photometric surveys is progressively available at
$z\sim 5$, based on more sophisticated selection criteria. In this
context, we present here an estimate of the space density of luminous QSOs
selected in a homogeneous way within the QUBRICS (QUasars as BRIght beacons
for Cosmology in the Southern hemisphere)
survey \citep{calderone19,Bou20}. This survey turned out to be
particularly efficient ($\sim 70\%$) and complete ($\sim 80\%$) in
finding very bright QSOs at high-z \citep{boutsia21}. This project,
joining together state-of-the-art surveys like SkyMapper, Gaia, and
WISE, is opening a new era of QSO searches at high-z. Adopting the
best-fit QSO luminosity function derived at $z\sim 4$ in our previous
work and applying the redshift evolution in the space density derived
from the present work, we improve the estimate of the AGN
photo-ionization rate at $z\sim 5$.

The structure of this paper is the following: in Section 2 we describe
the QUBRICS survey, the selection of $z\sim 5$ QSOs and their
completeness. In Section 3 we derive the QSO luminosity function at
$M_{1450}\sim -28.5$, discussing the evolution of the QSO space
density with redshift, and we compute the photo-ionization rate
produced by QSOs at $z\sim 5$. We discuss the reliability of these
results in Section 4, providing the concluding remarks and summary in
Section 5. Throughout the paper, we adopt the $\Lambda$ cold dark
matter ($\Lambda$-CDM) concordance cosmological model (H$_{0}$=70 km
s$^{-1}$ Mpc$^{-1}$, $\Omega_{M}$=0.3, and $\Omega_{\Lambda}$=0.7). All
magnitudes are in the AB photometric system.


\section{Data} \label{sec:selection}

\subsection{The QUBRICS Survey}

The QUBRICS Survey aims at finding the brightest high-z QSOs in the
Southern hemisphere \citep[][hereafter PaperI and PaperII]{calderone19,Bou20}
by adopting the Canonical Correlation Analysis (CCA)
machine learning techniques applied to wide multi-wavelength databases
(e.g. SkyMapper, Gaia, WISE). The first pilot exploratory observations
revealed relatively high efficiencies, of $\sim 80\%$. After many
observations, the QUBRICS Survey produced a list of more than 250 new
bright ($i_{psf}\le 18.0$) QSOs at $z>2.5$, assessing the success rate to
$\sim 70\%$ and the completeness to $\sim 80\%$.
The success rate of finding QSOs at $z>4.5$ for the QUBRICS survey is 1.6\%,
much lower than at $z>2.5$, since the original goal of our survey was to
find objects at $z\sim 3$. In the meanwhile, other
high-z bright QSOs are going to be discovered by QUBRICS, thanks to the
exploitation of new state-of-the-art machine learning techniques,
e.g. Probabilistic Random Forest \citep[PRF,][]{guarneri21}
or Extreme Gradient Boosting (XGB, Calderone et al. in prep.).

The new high-z QSOs discovered by the QUBRICS Survey are going to be
used to feed efficient high-resolution spectrographs in order to study
the properties of the IGM or to carry out tests of fundamental physics
(see e.g. PaperII for a discussion of the Sandage Test). Since the
QUBRICS project turns out to be an efficient survey, but also a
complete one, it has been used in \citet{boutsia21} to derive the
Luminosity Function of $z\sim 4$ QSOs at unprecedented high
luminosities ($M_{1450}\le -28.0$). In this paper we extend this
analysis at $4.5<z<5.0$, where information on the ultra-bright active SMBH
population is still scanty.

\subsection{Selection of ultra-bright QSOs at $z\sim 5$ in the QUBRICS Survey}

The main sample of QUBRICS described in PaperII has
been adopted in order to select spectroscopically confirmed QSOs at
$4.5\le z_{spec}\le 5.0$ with an i-band magnitude brighter than $i_{psf}=18.0$. 
Table \ref{tab:qsoz5} includes 14 QSOs that satisfy the above criteria.
The spectroscopic confirmation of these high-z QSOs have been carried out
by the QUBRICS team (PaperI, PaperII), as well as by other
independent observations \citep[e.g.][]{Veron10,Sch19a,Sch19b,Wolf20}.
It is worth noting that all these different groups carried out
independent surveys; nonetheless, several high-z QSOs in the southern
hemisphere have been recently confirmed in multiple surveys, as
indicated in the last column of Table \ref{tab:qsoz5}.

The absolute magnitudes at 1450 {\AA} rest frame wavelengths
($M_{1450}$) in Table \ref{tab:qsoz5} have been derived following
\citet{boutsia21}.
Starting from the apparent $i_{psf}$ magnitudes of SkyMapper and from 
the spectroscopic redshifts, the equation
\begin{equation}
M_{1450} = i_{psf}-5log(d_L)+5+2.5log(1+z_{spec})+K_{corr} \, ,
\end{equation}
has been adopted,
where $d_L$ is the luminosity distance expressed in parsec (pc). The
k-correction $K_{corr}$ is given by the expression
\begin{equation}
K_{corr}=-2.5\alpha_\nu log_{10}(\lambda_{obs}/(1+z_{spec})/\lambda_{rest}) \, ,
\end{equation}
where the adopted spectral slope of QSOs is $\alpha_\nu=-0.7$, 
$\lambda_{rest}=1450$ {\AA}, and $\lambda_{obs}=7799$ {\AA} is the central
wavelength of the $i_{psf}$ filter.

Alternatively, the absolute magnitudes $M_{1450}$ can be derived from the calibrated spectra,
as carried out by e.g. \citet{Glikman10}. In order to check the consistency of the $M_{1450}$
from imaging with that from spectroscopy, we carry out the calculation of the k-corrections
for the six QSOs in Table \ref{tab:qsoz5} observed by QUBRICS.
The difference between the two k-corrections is $-0.087\pm 0.020$. This
confirms that the absolute magnitudes derived from the $i_{psf}$ apparent
magnitudes are consistent with the one derived from the spectra at the $\lesssim$10\% level.
In the following, we will adopt the absolute magnitudes $M_{1450}$ derived from photometry.

\begin{table*}
\caption{The $i_{psf}\le 18.0$ QSO sample at $4.5\le z_{spec}\le 5.0$ in the QUBRICS Survey.}
\label{tab:qsoz5}
\begin{center}
\begin{tabular}{l c c c c c c c l}
\hline
$ID_{SkyMapper}$ & RA & Dec & z$_{spec}$ & $i_{psf}$ & $i^{err}_{psf}$ & M$_{1450}$ & $z_{\rm cca}$ & z$_{spec}$ References \\
DR1.1 & J2000 & J2000 & & AB & & \\
\hline
7596895   & 00:05:00.19 & -18:57:15.43 & 4.560 & 17.82 & 0.03 & -28.450 & 3.64 & Wolf20 \\
317802750 & 00:12:24.99 & -48:48:29.86 & 4.621 & 17.72 & 0.17 & -28.588 & 5.44 & PaperII,Wolf20 \\
7869715   & 01:35:39.28 & -21:26:28.42 & 4.940 & 17.99 & 0.08 & -28.460 & 4.77 & Wolf20 \\
318204033 & 03:07:22.89 & -49:45:48.24 & 4.728 & 17.41 & 0.04 & -28.974 & 4.87 & P01,Veron10 \\
10431842  & 03:55:04.85 & -38:11:42.41 & 4.545 & 17.82 & 0.16 & -28.467 & 4.42 & P01,Veron10 \\
54680559  & 11:10:54.68 & -30:11:29.88 & 4.779 & 17.35 & 0.03 & -29.035 & 4.59 & PaperII,PSELQS,Wolf20 \\
65100743  & 12:05:23.13 & -07:42:32.65 & 4.690 & 17.88 & 0.03 & -28.455 & 5.18 & R06,Veron10 \\
3436512   & 21:11:05.60 & -01:56:04.14 & 4.891 & 17.91 & 0.02 & -28.526 & 5.56 & PaperI,Wolf20 \\
304245360 & 21:19:20.85 & -77:22:53.17 & 4.558 & 17.86 & 0.07 & -28.412 & 4.93 & PaperI,Wolf20 \\
397340    & 21:57:28.21 & -36:02:15.11 & 4.771 & 17.37 & 0.02 & -29.009 & 5.05 & PaperII,Wolf20 \\
4368005   & 22:21:52.88 & -18:26:02.87 & 4.520 & 17.87 & 0.08 & -28.379 & 3.11 & PSELQS \\
5392050   & 22:39:53.67 & -05:52:19.81 & 4.558 & 17.95 & 0.09 & -28.321 & 1.58 & SL96,Veron10 \\
1913850   & 23:04:29.88 & -31:34:27.02 & 4.840 & 17.77 & 0.04 & -28.636 & 4.69 & Wolf20 \\
308375290 & 23:35:05.86 & -59:01:03.33 & 4.540 & 17.60 & 0.02 & -28.664 & 5.36 & PaperII,Wolf20 \\
\hline
\end{tabular}
\tablecomments{
      \\
      The references for z$_{spec}$ are: \\
      SL96 refers to \citet{StoLom1996}. \\
      P01 refers to \citet{peroux01}. \\
      R06 refers to \citet{Riechers06}. \\
    Veron10 refers to \citet{Veron10}. \\
      PSELQS refers to \citet{Sch19b}. \\
      Wolf20 refers to \citet{Wolf20}. \\
      }
\end{center}
\end{table*}

\subsection{Completeness of $z\sim 5$ QSO Sample}

In order to estimate the completeness of the QSO sample
in Table \ref{tab:qsoz5}, it is useful to
consider the results obtained by similar surveys. \citet{Wolf20}
searched for bright QSOs at $z>4.5$ using data
from SkyMapper, Gaia, and WISE. They found 23 QSOs at $4.5\le z_{spec}\le 5.0$
and $M_{1450}\le -28.0$ in 12,500 sq. deg. in the southern hemisphere.
Since their survey area is similar to the QUBRICS one (12,400 sq. deg.),
and their magnitude cut is similar to the one adopted here, it is surprisingly
that they retrieve 65\% more QSOs than in our sample. Checking carefully
Table \ref{tab:qsoz5}, however, it is evident that our sample is limited to
slightly brighter magnitudes ($M_{1450}\le -28.3$) than their limit.
If the \citet{Wolf20} sample is restricted to $M_{1450}\le -28.3$, 19
QSOs are left, reducing the discrepancy with our sample within the mutual
statistical uncertainties.

Of the 23 QSOs by \citet{Wolf20} with $4.5<z<5.0$ and $M_{1450}\le
-28.0$, only 12 objects are part of our QUBRICS sample. Nine QSOs of
\citet{Wolf20} have an i-band magnitude in SkyMapper DR1.1 fainter
than 18.0, so out of the QUBRICS selection criteria. Two QSOs of
\citet{Wolf20} have $i_{psf}<18.0$ but they are not part of our
sample. For one of these two QSOs (J151443.82-325024.8) the galactic latitude is
below the adopted threshold in PaperI and PaperII, i.e. $|g_{lat}|\ge
25^{deg}$, and thus it does not belong to our main sample.
Only one QSO, J145147.04-151220.1 at $z_{spec}=4.76$, has
a $i_{psf}=17.1826$ in SkyMapper DR2, but it is not extracted by
SkyMapper DR1.1. It is worth noting that this is a very bright QSO
($M_{1450}=-29.29$ in Wolf et al. 2020), and it is missing from our
sample since the QUBRICS survey has been based on SkyMapper DR1.1.
Based on this comparison, a simple determination of
the completeness of the QUBRICS sample can be estimated of the order of
92.3\% (12/13). Two objects in our sample are not present in \citet{Wolf20},
indicating that the completeness level of their survey is $\sim 90\%$.
It is worth noting that the completeness correction for bright QSOs at $z\sim 4$
in the QUBRICS survey is slightly lower, of $\sim 84\%$,
as discussed in \citet{boutsia21}.
Recently, \citet{Onken21} have released an incremental sample of
119 new QSOs at $z>4$. We have checked that they are all fainter than our
magnitude limit of $i_{psf}=18.0$, so it is not possible to use their results
for a refinement of the completeness calculation for the QUBRICS survey.

It is possible to estimate the incompleteness in an alternative way,
as discussed in \citet{boutsia21}. The main sample of QUBRICS has been
extracted from the SkyMapper DR1.1 database, but it is possible that
this survey is not complete. We have used the Dark Energy Survey
\citep[DES,][]{des21} to select all the known QSOs at $4.5<z<5.0$ and
$M_{1450}\le -28.0$. We have found 5 QSOs in $\sim 5000$ sq. deg. area
satisfying these criteria. Of these 5 QSOs, only 4 objects have an
i-band magnitude determination in SkyMapper DR1.1, indicating that the
completeness of the latter is 80\% (4/5). Given the low number
statistics of the QSO sample based on DES, of only 5 objects, however,
we prefer not to use this completeness in the following analysis.

Following \citet{boutsia21}, another source of incompleteness is the
fraction of candidates still awaiting for spectroscopic confirmations.
Starting from the QUBRICS main sample in PaperII, there are 7 objects
with a photometric redshift (based on CCA) in the range
$4.5<z<5.0$. Out of these 7 candidates, 5 objects have been confirmed
to be QSOs at $4.5\le z_{spec}\le 5.0$, and they are included in Table
\ref{tab:qsoz5}. Two candidates do not have yet spectroscopic
identification, and future observations could increase the number of
bright QSOs at $z\sim 4.8$.  One of these two candidates, however, is
slightly extended on the i-band of SkyMapper, and for this reason it
could be a bona-fide low redshift object. In addition, there are
three new QSO candidates at $z>4.5$ selected with XGB and photometric
redshifts, as will be described in future papers by this
collaboration. We decide to not include in our completeness
correction the term owing to the spectroscopic missing sample, but we
are aware that the number of ultra-bright QSOs at $z\sim 5$ could
significantly increase in the future.

Based on the above considerations, we have decided to adopt as
completeness factor for our sample the conservative 92.3\% level
derived above. We should take into account, however, that this
estimate could be slightly lower (completeness of the order of
$\sim 70-80\%$), as derived by the DES survey or taking into
account the number of
candidates still awaiting for spectroscopic identifications.
In this case, the QSO space density at $z\sim 5$ would be even higher than the
present estimates. Future spectroscopic follow-up observations would reduce
the uncertainties on the completeness correction factor.


\section{Determination of the space density of ultra-bright QSOs
at $z\sim 5$} \label{sec:LFunction}

We have derived the co-moving space density of $4.5<z<5.0$ QSOs at
$M_{1450}\lesssim -28$ by adopting the standard $1/V_{max}$ approach, as
described in detail in \citet{boutsia21}. Due to the low number of
bright QSOs in the QUBRICS survey, we decided to compute the space
density in an absolute magnitude interval of $-29.3\le M_{1450}\le
-28.3$, that includes all the 14 QSOs in Table \ref{tab:qsoz5}. The
adopted lower bound in luminosity ($M_{1450}=-28.3$) corresponds to
the magnitude limit of the survey in the i-band $i_{psf}\le 18.0$ at
$z\sim 4.5$.

We have corrected the observed space density for incompleteness by
adopting the correction factor of 92.3\%, as discussed above. The
derived space density $\Phi$ is associated to an absolute magnitude of
$M_{1450}=-28.6$, the mean value of the QSOs in the ultra-bright
bin. Errors on $\Phi$ have been derived by adopting the
Poissonian statistics, since $N_{QSO}>10$. Table \ref{tab:lfobs}
summarizes the resulting space density $\Phi$ obtained from the
QUBRICS survey.

\begin{table*}
\caption{The $1/V_{max}$ space density $\Phi$ of $4.5\le z\le 5.0$ QSOs
in the QUBRICS footprint.}
\label{tab:lfobs}
\begin{center}
\begin{tabular}{c c c c c c}
\hline
\hline
Interval & $<M_{1450}>$ & $N_{QSO}$ & $\Phi$ & $\sigma_\Phi(up)$
& $\sigma_\Phi(low)$ \\
 & & & $cMpc^{-3}$ & $cMpc^{-3}$ & $cMpc^{-3}$ \\
\hline
$-29.30\le M_{1450}\le -28.30$ & -28.60 & 14 & $3.115\times 10^{-10}$ & $1.077\times 10^{-10}$ & $8.251\times 10^{-11}$ \\
\hline
\hline
\end{tabular}
\end{center}
The space density $\Phi$ has been corrected for incompleteness,
as discussed in the main text. A completeness correction factor
of 92.3\% has been adopted in this case.
\end{table*}

\begin{figure*}[h]
\plotone{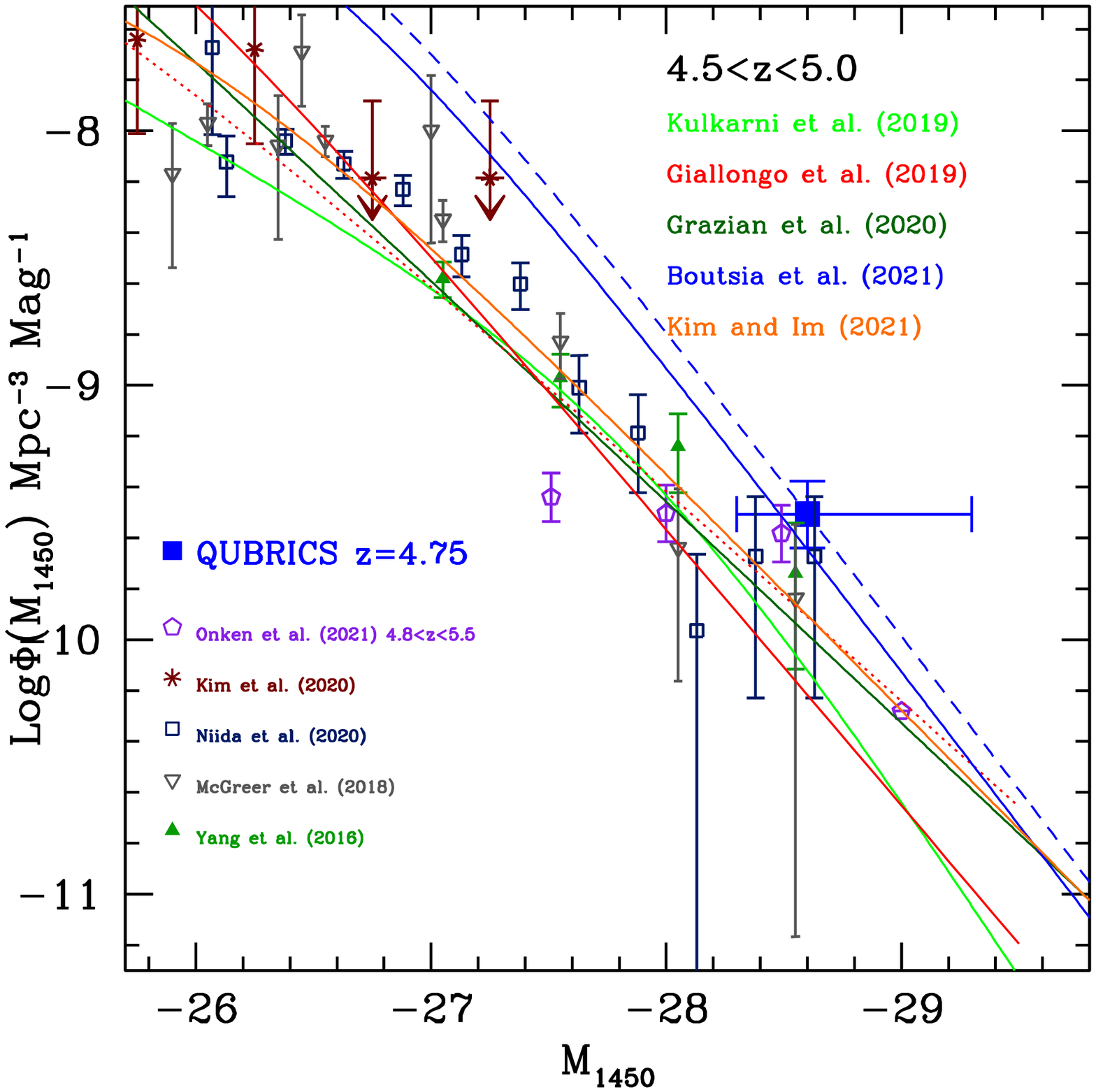}
\caption{The luminosity function of QSOs at $4.5\le z_{spec}\le 5.0$ from
QUBRICS (blue filled squares) compared to other luminosity functions
from the recent literature. All the data points and curves have been
shifted to $z=4.75$ adopting a pure density evolution with $\gamma=-0.25$,
as found in this work. The blue square indicates the mean absolute magnitude $<M_{1450}>$
of the QSOs inside the bin, marked by the blue horizontal bar. The blue continuous line
is not the fit to the data points, but it is the best fit of the QSO
luminosity function at $z\sim 4$ by \citet{boutsia21}, evolved at
$z=4.75$ with $\gamma=-0.25$, as well.
The blue dashed line is the luminosity function of \citet{boutsia21},
with a normalization factor that allows it to overlap with the
observed non-parametric point from QUBRICS. The red dotted line is the
model 3 of \citet{giallongo19}, while the red continuous line is their
model 4.}
\label{fig:lfz5}
\end{figure*}

The blue filled square in Fig. \ref{fig:lfz5} shows the space density
$\Phi$ of $4.5\le z\le 5.0$ QSOs in the QUBRICS footprint. For comparison,
we have plotted in the same figure the determination of the QSO luminosity
function by recent works at $z\sim 4-5$. When the mean redshift of these
surveys is different from $z=4.75$, their data points and curves have been
shifted by adopting the pure density evolution recipe with $\gamma=-0.25$, as
we will discuss in detail in Section \ref{sec:lfevol}.
The green, red, dark-green, and orange curves in Fig. \ref{fig:lfz5}
show the best-fit luminosity functions of
\citet{kulkarni19,giallongo19,grazian20,kimim21}.
The blue continuous line in Fig. \ref{fig:lfz5} is not the best fit to
the QUBRICS data point, but it is the luminosity function of
\citet{boutsia21} derived at $z\sim 3.9$ and evolved at $z=4.75$ by
adopting a pure density evolution recipe with $\gamma=-0.25$.
The blue dashed line in Fig. \ref{fig:lfz5} is the same luminosity
function as above, but with an {\em ad hoc} normalization, chosen to
fit the QUBRICS data point at $z\sim 5$.

At these very bright luminosities ($M_{1450}\le -28.5$),
our space density is a factor of $\sim 3$
higher than previous determinations by \citet{Yang16,McGreer18,Niida20}
or than the best fit by \citet{kulkarni19,giallongo19,grazian20,kimim21}.
Since our luminosity function has been derived simply by dividing the
number of confirmed QSOs at $z\sim 5$ by the cosmological volume of
the Universe corresponding to $4.5<z<5.0$ and 12,400 sq. deg. area,
with a small correction for completeness of 92.3\%, it is quite
surprising that $\Phi$ is significantly higher than literature
estimations at the same redshifts. A possible explanation could be that the
latter are heavily affected by strong incompleteness of a factor of $\sim 3$.
This is not surprising, however, given that it has been shown recently that
even at $z\sim 4$ the previous determination of the QSO luminosity functions
were incomplete by at least $\sim 30-40\%$ \citep{Sch19a,Sch19b,boutsia21}.
A similar result at $M_{1450}\sim -28.5$ has been obtained by \citet{Onken21}
on a partially overlapping sky area.
They find that the space density of $z\sim 5$ ultra-bright QSOs
is 3 times higher than
previous determination by the Sloan Digital Sky Survey (SDSS),
and it is similar to the one derived
by the QUBRICS survey. This is a clear
indication that the luminosity function from our sample is in
agreement with these recent estimates, indicating that previous
determinations in the past were possibly affected by severe incompleteness.

Summarizing, a first result, drawn from Fig. \ref{fig:lfz5}, is that the
space density of ultra-bright QSOs at $z\sim 5$ is significantly higher,
at least by a factor of 3, from previous determinations in the past, while it is
fully consistent with the recent results by \citet{Wolf20,Onken21}.

\subsection{The slope of the bright end of the QSO Luminosity
Function at $z\sim 5$} \label{sec:slope}

The non-parametric space density derived in Fig. \ref{fig:lfz5} has
limited information on the shape of the QSO Luminosity Function at
$z\sim 5$. Despite the low number of QSOs in the QUBRICS survey and
the relatively small range in luminosity covered by our sample, we try to
constrain the bright-end slope by adopting the Maximum Likelihood
formalism by \cite{Marshall83}. Following \citet{boutsia21}, we adopt
a double power-law parameterization for the Luminosity Function:

\begin{equation}
\phi = \frac{\phi^*}{10^{0.4(M-M^{*}_{1450})(\alpha+1)}+10^{0.4(M-M^{*}_{1450})(\beta+1)}} \,,
\end{equation}

where $\alpha$ and $\beta$ are the faint and bright-end slopes of the
luminosity function, $M^{*}_{1450}$ is the absolute magnitude of the break
and $\phi^*$ is its normalization (which is not a free parameter of the Maximum
Likelihood calculations). Fixing the knee of the luminosity
function and the faint-end slope to $M^{*}_{1450}=-26.50$ and $\alpha=-1.85$,
from the $z\sim 3.9$ parameterization provided by \citet{boutsia21}, we obtain a
best fit to the bright-end slope of $\beta=-4.64$, with a range between
-6.19 and -3.43 at 68\% confidence level probability. These values do not
change if we fix the break to -26.0, since the QUBRICS data are limited
to $M_{1450}\le -28.30$, and consequently they cannot provide any leverage
on the break of the QSO luminosity function. If we leave as free parameters
both the bright-end slope and the break, we obtain a 68\% confidence
interval of $\beta\le -3.03$ and $M^{*}_{1450}\ge -29.80$. This indicates
again that the break $M^{*}_{1450}$ is practically unconstrained and the
bright-end slope is relatively steep.
If we divide our sample in two absolute magnitude intervals
$-29.30<M_{1450}<-28.80$ and $-28.80<M_{1450}<-28.30$ and compute the 
non-parametric $1/V_{max}$, then the resulting bright-end slope derived 
from these two points is $\beta=-4.48$, consistent with the previous values.
The bright-end slope is still ripid ($\beta=-4.97$) if we divide
our sample in two
uneven bins, containing 7 objects each, i.e. $-29.30<M_{1450}<-28.50$ and
$-28.50<M_{1450}<-28.30$. This is a starting indication that
the bright-end slope of $z\sim 5$ QSO Luminosity Function is compatible
with $\beta\sim -4$, as been derived at lower redshifts by
\citet{Sch19a,Sch19b,boutsia21} and with $\beta=-3.84$ derived
at $z\sim 5$ by \citet{Onken21}. We do not find an indication of a 
flattening of the bright-end slope of the QSO Luminosity Function going at high
redshifts, as previously suggested by
\citet{Fan2001,fan06,jiang08,willott10,masters12}.

\subsection{The evolution of the QSO space density with redshift}
\label{sec:lfevol}

Extrapolations of the blue lines in Fig. \ref{fig:lfz5} at absolute
magnitudes fainter than $M_{1450}\sim -28$ are a factor of $\sim 3-5$
above the previous determination of the QSO space densities at $z\sim
5$ by e.g. \citet{McGreer18,Yang16,Niida20,Kim20}. Since the QUBRICS
survey is typically sensible only to ultra-bright QSOs at these
redshifts, it is worth asking whether this dramatic disagreement is
due to a strong incompleteness of the previous surveys at
$M_{1450}\sim -26$ or, alternatively, if the extrapolations of the
blue lines in Fig. \ref{fig:lfz5} to the faint end are too optimistic
and the luminosity function of quasars near the break is indeed far lower. In
order to provide a plausible answer to this question, we compare here
the space densities of bright QSOs and faint AGNs at $z>2.5$.

\begin{figure}
\plotone{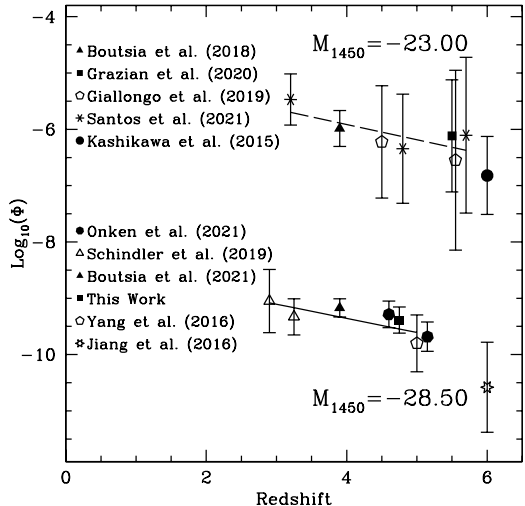}
\caption{The space density of bright QSOs at $M_{1450}=-28.5$ and faint AGNs
at $M_{1450}=-23.0$ at $z>2.5$ from this work and recent surveys.}
\label{fig:rhoz}
\end{figure}

Fig. \ref{fig:rhoz} (bottom part) shows the space density of bright QSOs at
$M_{1450}=-28.50$ as a function of the mean redshifts of the surveys
at $z>2.5$ collected from the literature. Fitting only the $z<=5.0$
data, obtained by \citet{Sch19b} at $z=2.90$ and 3.25, by the QUBRICS
survey at $z=3.9$ by \citet{boutsia21}, at $z=4.75$ (this work) and at
$z=5.0$ by \citet{Yang16}, we obtain a best-fit parameter of
$\gamma=-0.25$, by adopting an evolution of the space density as
$\Phi(z)=\Phi(z=4.0)\cdot 10^{\gamma\cdot (z-4.0)}$. The continuous line in
Fig. \ref{fig:rhoz} indicates the resulting best fit to the $z\le 5.0$
data points at $M_{1450}=-28.50$. More interestingly, neglecting the z=5.0
point by \citet{Yang16}, the evolution is even milder with
$\gamma=-0.11$. The data point at $M_{1450}=-28.50$ of
\citet{Onken21} has been added for comparison in Fig. \ref{fig:rhoz},
but it cannot be used
during the fitting procedure, since it is not independent from the
space density in QUBRICS.
It shows however a good agreement with the best fit of $\gamma=-0.25$.

If we use all the data points from $z=2.9$ to $z=6.0$,
then the resulting best-fit parameter is much steeper, $\gamma=-0.69$,
indicating an accelerated evolution of the space density of bright
QSOs at $z>5$. If we limit the analysis to $z\ge 5.0$, then the
redshift evolution of $\Phi(z)$ is even more pronounced, with
$\gamma=-0.78$. Interestingly, limiting the redshift to $z\ge 4.5$ the
evolution is even more dramatic, with a best fit of
$\gamma=-0.91$. These latter values are compatible with the
independent estimates carried out in the past by \citet{Ross13}
($\gamma=-0.809$) and by \citet{Yang16} ($\gamma=-0.81$).

We have collected on the top of Fig. \ref{fig:rhoz} the space
densities $\Phi$ at a fainter absolute magnitude of $M_{1450}=-23.0$,
in order to check whether the redshift evolution of the faint side of
the AGN Luminosity Function is similar to the bright one. Of course,
the statistics of faint AGNs at very high redshifts mainly rely on
photometric redshifts derived from multi-wavelength data and often
lack spectroscopic confirmation. In these cases the X-ray detection is
a valuable criterion to include a Lyman break galaxy or a
Lyman-$\alpha$ emitter in the high redshift AGN population
\citep[e.g.][]{fiore12,giallongo15,boutsia18,giallongo19,grazian20}. We
have included in Fig. \ref{fig:rhoz} the space densities derived from
our CANDELS and COSMOS surveys. We have also included the recent
results obtained by the SC4K survey \citep{santos21} at $z=3.1\pm
0.4$, $z=4.7\pm 0.2$, and $z=5.4\pm 0.5$. The latter is based on 12
medium band and 4 narrow band filters used to select LAEs at various
redshifts. The AGN fraction within the LAE sample has been selected
by means of strong X-ray or Radio detection, as detailed in
\citet{santos21}. It is worth noting the remarkable agreement of the
space densities derived at about the same redshifts by our faint
surveys, mainly based on broad band photometry, and the CS4K survey,
based on narrow band photometry where the redshift accuracy is higher,
thanks to the emission line detection. Finally, we have included the
faint color-selected AGN sample by \citet{kashikawa15} at
$z=6.0$. Although limited by a simple and conservative color
selection, it allows an extension of the analysis to the highest
redshifts available for a faint AGN sample.

At $3\le z\le 5.5$, the best-fit parameter of the redshift evolution
at $M_{1450}=-23.0$ is
$\gamma=-0.27$, comparable with the one at brighter luminosities.
The dashed line in Fig. \ref{fig:rhoz} (top) indicates the
resulting best fit to the $z\le 5.5$ data at $M_{1450}=-23.0$.
If we include the data by \citet{kashikawa15} at $z=6.0$, the slope is
slightly steeper, with $\gamma=-0.33$, but not comparable to the rapid
evolution at $z>5$ found for brighter QSOs, as discussed above.

We can draw two main considerations from Fig. \ref{fig:rhoz}:
\begin{enumerate}
\item
the space densities of bright QSOs and faint AGNs at $3\le z\le 5$
show a similar evolution in redshift, indicating that the QSO luminosity
function plausibly follows a pure density evolution, i.e. with a rigid shift
in its normalization.
\item
The density evolution at $3\le z\le 5$ is milder than previous
determinations, with a best-fit parameter of $\gamma=-0.25$.
\end{enumerate}

The data at $z>5.5$ in Fig. \ref{fig:rhoz} at bright luminosities
seem to be compatible with an accelerated
evolution of the space density of luminous QSOs with respect to $z<5.5$.
Its is possible, however, that the different density evolution shown in
Fig. \ref{fig:rhoz} is due to an incompleteness of the surveys at
$z\gtrsim 5$ or due to an underestimation of the incompleteness factors.
Recent surveys \citep{Sch19b,boutsia21},
indeed, have shown that the first attempts to measure the space
densities of bright QSOs at high-z by the SDSS could possibly suffer from
incompleteness by a large amount of $\sim 30-40\%$ at $z\sim 4$,
reaching a factor of
3 incompleteness at $z\sim 5$ \citep{Wolf20,Onken21}. If such an
incompleteness is affecting also the data points at $z>5$, and plausibly the
incompleteness is stronger at $z\sim 6$ than at $z\sim 5$, then the
observed drop could only be due to a spurious trend.
Future investigations on the
number densities of very bright QSOs at $z>5$ will confirm or reject
this hypothesis.

\subsection{The ionizing background produced by QSOs at $z\sim 5$}

The photo-ionization rate $\Gamma_{HI}$ produced by bright QSOs and
faint AGNs at $z\sim 5$ is derived here. We start from the recent
determination of the $z\sim 4$ QSO luminosity function by
\citet[][see their Table 4]{boutsia21}, rescaled to $z=4.75$ with a
pure density evolution, parameterized by the newly determined
coefficient $\gamma=-0.25$, as discussed in the previous sub-section.

\begin{figure}
\plotone{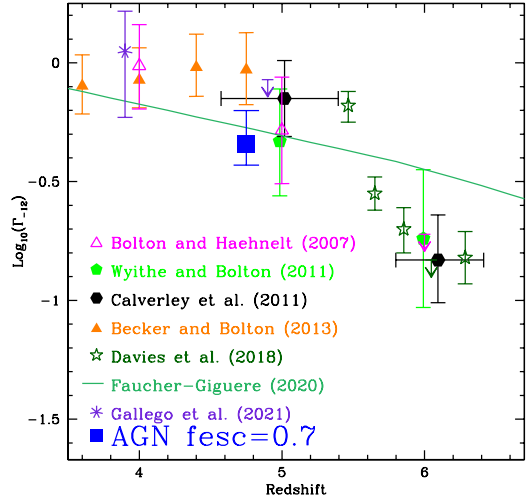}
\caption{The photo-ionization rate $\Gamma_{HI}$ (in units of $10^{-12} s^{-1}$)
produced by the AGN population at $z\sim 5$ (blue square), assuming
a Lyman Continuum escape fraction of 70\% and a pure density evolution of
the QSO luminosity function with $\gamma=-0.25$.
A collection of the photo-ionization
rates from different methods (e.g. Lyman-$\alpha$ forest fitting,
Proximity effect,
QSO near zone size) has been carried out for comparison.
The green continuous line
is the prediction of the model by \citet[][]{Faucher20}.}
\label{fig:gamma}
\end{figure}

Following the calculations carried out by \citet{boutsia21}, we first
derived the luminosity density at 1450 {\AA} rest frame by integrating
the QSO luminosity function, multiplied by the luminosity, from
$M_{1450}=-30.0$ down to $M_{1450}=-18.0$. The exact value of the
brighter integration limit does not significantly influence the total
amount of the emissivity, since the bulk of UV photons are produced by
AGNs with luminosity close to $L^*$, while the rare, very bright QSOs
are contributing only to few percents of the total emissivity
\citep{giallongo19}. We have assumed an escape fraction of Lyman
continuum (LyC) photons of $70\%$, in agreement with recent results by
\citet{cristiani16,grazian18,romano19} and a spectral slope of
$\alpha_\nu=-0.61$ at $\lambda>912$ {\AA} rest frame and
$\alpha_\nu=-1.7$ at $\lambda\le 912$ {\AA} rest frame, in agreement
with \citet{lusso15}. The MFP of HI ionizing photons at $z=4.75$ is 17.4 proper
Mpc, following the relation by \citet{worseck14}. We consider also the
factor of 1.2 due to radiative recombination in the IGM
\citep{DAloisio18}. Using the same formalism adopted by
\citet{lusso15}, we obtain, from the AGN emissivity, a photo-ionization
rate at $z=4.75$ of $\Gamma_{HI}=0.46^{+0.17}_{-0.09}\times 10^{-12} s^{-1}$,
shown as a blue square in Fig. \ref{fig:gamma}.

Comparing this value with the recent estimates of the ionizing UV
background at these redshifts summarized in Table \ref{tab:gammaz5}
and shown in Fig. \ref{fig:gamma},
it turns out that $z\sim 5$ AGNs are able to produce $\sim 50-90\%$ of the LyC
photons. This fraction can reach $\sim 100\%$ if we consider
the uncertainties on the
QSO contribution to $\Gamma_{HI}$ and the variance on the
determination of the photo-ionization rate measured by the Lyman
forest or by the proximity effects
\citep[e.g.][]{wyithe11,calverley11,davies18}. Thus it emerges from
these calculations that the contribution by $z\sim 5$ AGNs to the
ionizing background is probably not negligible.
In the next section we will discuss the implications of the LyC escape fraction
and the faint-end slope of the QSO luminosity function on the estimate of the
photo-ionization rate produced by high-z AGNs.

\begin{table*}
\caption{The photo-ionization rate $\Gamma_{HI}$ at $z\sim 5$.}
\label{tab:gammaz5}
\begin{center}
\begin{tabular}{c c c c c c}
\hline
\hline
Reference & Method & redshift & $\Gamma_{HI}$ & $+\sigma_{\Gamma_{HI}}$ &
$-\sigma_{\Gamma_{HI}}$ \\
 & & & in $10^{-12} s^{-1}$ & & \\
\hline
THIS WORK ($\gamma=-0.25$) & AGN & 4.75 & 0.46 & +0.17 & -0.09 \\
\hline
\citet{Bolton07} & Ly-$\alpha$ forest & 5.00 & 0.52 & +0.35 & -0.21 \\
\citet{wyithe11} & QSO near-zone size & 4.985 & 0.47 & +0.31 & -0.19 \\
\citet{calverley11} & Proximity effect & 5.02 & 0.71 & +0.32 & -0.22 \\
\citet{BeckerBolton13} & Ly-$\alpha$ forest & 4.75 & 0.94 & +0.40 & -0.27 \\
\citet{Faucher20} & Model & 4.80 & 0.525 & \nodata & \nodata \\
\citet{Gallego21} & Fluorescent Lyman-$\alpha$ & 4.9 & $<$0.85 & \nodata & \nodata \\
\hline
\end{tabular}
\end{center}
The value of the ionizing background produced by AGNs has been
estimated through the best-fit luminosity function of
\citet{boutsia21} evolved at $z=4.75$ with a pure density evolution
with $\gamma=-0.25$.
\end{table*}

\section{Discussion}

\subsection{The Reliability of the QUBRICS QSO luminosity function at $z\sim 5$}

The space density of luminous QSOs at $z>4.5$ of the QUBRICS survey
shown in Fig. \ref{fig:lfz5} is a factor of $\sim 3$ higher than previous
estimates. Since it is based on a wide area, covering approximately
one third of the whole sky, it is implausible that cosmic variance
plays a role in over-estimating the bright QSO counts. A similar result has been
obtained also by \citet{Onken21} at $z>4.4$ in a sky area partially overlapping
with the QUBRICS survey. The reason for the huge discrepancies with respect
to previous surveys is probably due to more efficient and complete selection
criteria on the recent surveys, as shown by \citet{Sch19b} at $z>3$ or by
\citet{boutsia21} at $z\sim 4$.

There are indications, on the contrary, that the present determination of the
space density could be even higher for a number of reasons:
\begin{enumerate}
\item
the colors of $z\sim 5$ QSOs are similar to late-type stars.
In order to gain efficiency in the selection of QSO candidates for
spectroscopic confirmations, stringent color criteria have been
adopted \citep[e.g.][]{McGreer18,matsuoka18}. A price to be paid in order
to have high spectroscopic efficiency is the low completeness of the
adopted selection criteria.
\item
Several high-quality candidates based on colors or CCA criterion
still need to be
spectroscopically identified yet. Two bona-fide QSO candidates with $z_{CCA}>4.5$
and three high-quality candidates with $z_{XGB}>4.5$ or selected with PRF
have not been spectroscopically observed yet.
\item
The space density shown in Fig. \ref{fig:lfz5} is based on the CCA
selection criterion \citep{calderone19,Bou20}, which has not been
tuned to be particular complete at $z>4.5$, but has been design to have
high efficiency in a broader redshift range, i.e. at $z>2.5$.
The CCA selection indeed recovers only 5 of the 14 QSOs
known and discovered in the QUBRICS footprint (see Table \ref{tab:qsoz5}).
The QUBRICS team is currently exploring different selection criteria, e.g.
the Probabilistic Random Forest \citep[PRF,][]{guarneri21} and the Extreme
Gradient Boosting (XGB, Calderone et al. in prep.), which could be tuned
in order to have high completeness at $z>4.5$.
\end{enumerate}

Based on the above considerations, it is easy to conclude that the QSO
luminosity function shown in Fig. \ref{fig:lfz5} is not the final
measurement at $z>4.5$ and at bright UV magnitudes ($M_{1450}\lesssim -28$).
Recent attempts to find bright QSOs at $z>4.5$ are under way
\citep[e.g.][]{Wolf20,Wenzl21,Onken21} and it is possible that additional
$z\sim 5$ QSOs will be found in the near future in the QUBRICS footprint.

\subsection{Uncertainties in the estimate of the Photo-ionization rate at $z\sim 5$}

The Photo-ionization rate produced by bright QSOs and faint AGNs at
$z=4.75$ is relatively high if the shape of the QSO luminosity
function does not change from z=3.9 to $z\sim 5$. A rigid shift of the
QSO Luminosity Function
with a pure density evolution seems to be confirmed by the data shown in
Fig. \ref{fig:rhoz}, where it is clear that the redshift evolution is
similar both for bright QSOs and for faint AGNs. The new
determinations of $\Phi$ by \citet{Onken21} at $4.4<z<4.8$
and $4.8<z<5.5$ at the
absolute magnitude $M_{1450}=-28.5$ are in agreement with the redshift
evolution shown in Fig. \ref{fig:rhoz}.

A pure density evolution of the AGN luminosity function at $z>3$, however,
is not in full agreement with the results of recent works
which find a flatter slope for the faint end of the QSO luminosity
function at high-z \citep[e.g.][]{matsuoka18,kulkarni19,kimim21}.
If the faint end of the $z\sim 5$ luminosity function is flatter than our extrapolations, the
contribution of high-z AGNs to the ionizing background will be lower
than our estimate in Table \ref{tab:gammaz5} and Fig. \ref{fig:gamma}.

In order to provide a quantitative estimate of the photo-ionization rate in the case of
a flatter faint-end luminosity function, we consider the pure density evolution provided by
\citet{kimim21}. According to their cases 1, 2, and 3, the AGN luminosity function at $z=4.75$
is parameterized by a two-power law with slope $\alpha\sim -1$ and low space densities at $M_{1450}>-24$. Adopting their parameterizations
for the luminosity function at $z=4.75$, we obtain a photo-ionization rate of
$\Gamma_{HI}=0.058-0.063\times 10^{-12} s^{-1}$. It corresponds to 6-13\% of the HI ionizing 
background measured at $z\sim 5$ (see Table \ref{tab:gammaz5}). Thus, in case of a flat
faint-end luminosity function at $M_{1450}>-24$,
the contribution of QSOs and AGNs to the photo-ionization
rate at $z\sim 5$ is not dominant.

It is worth noting, however, that the flat luminosity function by \citet{kimim21}
is in tension with the luminosity functions of \citet{glikman11} and
\citet{boutsia18} at $z\sim 4$ and with the ones by \citet{giallongo19} and
\citet{grazian20} at $z\sim 5.5$. Moreover, a flat luminosity function at $z\sim 5$
seems not fully compatible with the redshift evolution of the observed space densities,
as summarized in Fig. \ref{fig:rhoz}.

The determination of the photo-ionization rate by QSOs also depends on
the Lyman continuum escape fraction of the AGN population. There are
indications that this parameter is around 70\% at $z\sim 4$ and it is
almost constant both in luminosity \citep{grazian18} and in redshift
\citep{cristiani16,romano19}. Stacking thousands of SDSS QSOs at $3.5\lesssim z\lesssim 5.5$,
\citet{prochaska09} and \citet{worseck14} derived the MFP of HI ionizing radiation,
assuming that the LyC escape fraction of these QSOs is close to 100\%. The fact that
their MFPs are in agreement with the results of \citet{inoue14}, based on the
statistics of intergalactic absorbers, seems to indicate that $f_{esc}(LyC)\sim 100\%$,
at least for very bright QSOs ($M_{1450}<-27$). For these reasons, we have assumed
$f_{esc}(LyC)=70\%$ for the entire luminosity regime where the
QSO luminosity function has been integrated to compute the emissivity
by active SMBHs.

High values of the escape fraction for high-z AGNs have been questioned by
\citet{micheva17}, who found $f_{esc}(LyC)\sim 30-50\%$ at $z\sim 3$ based on
deep narrow-band photometry at $\lambda\sim 3600$ {\AA}.
Recently, \citet{iwata21} have obtained an $f_{esc}=0.36\pm 0.10$ for $M_{1450}\le -24$
AGNs at $z\sim 3.5$ and an $f_{esc}=0.25\pm 0.10$ for $M_{1450}>-24$.
If we adopt the AGN luminosity function of \citet{boutsia21}, evolved at $z=4.75$
with a pure density evolution with $\gamma=-0.25$, and we consider the escape fraction by
\citet{iwata21}, we obtain a photo-ionization rate of $\Gamma_{HI}=0.19\times 10^{-12} s^{-1}$,
which is 21-41\% of the HI ionizing background provided in Table \ref{tab:gammaz5}.
Thus, the contribution of high-z AGNs to the ionizing background still remains
significant although not dominant, in the case of a low LyC escape fraction for AGNs.

It is possible, however, to reconcile the results of \citet{micheva17} and \citet{iwata21} with the
one by \citet{cristiani16,romano19}. The latter indeed measured the LyC escape
fraction for $>$2000 QSOs at $z\ge 3.6$ at an absolute magnitude
brighter than $M_{1450}\sim -27$, while the former adopted a sample
of 94 AGNs at $3<z<4$ and $-26.5\le M_{1450}\le -19.0$, with the bulk
around $M_{1450}\sim -24$. Thus, the LyC escape fraction could show a mild dependency on the
absolute magnitudes of the AGNs, going from $>70\%$ at $M_{1450}\sim -28$ down to
$\sim 36\%$ at $M_{1450}\lesssim -24$ and $\sim 25\%$ at $M_{1450}> -24$, as found by \citet{iwata21}.
It is worth noting, as discussed in \citet{giallongo19} and \citet{boutsia21},
that the bulk of ionizing photons is produced by
$L\ge L^*$ AGNs. If the escape fraction of faint AGNs is significantly lower than 70\% at
$M_{1450}\gtrsim -23$, we expect that the total ionizing
radiation produced by the whole AGN population remains substantially
high, close to 50-100\% of the
photo-ionization rate. Based on these considerations,
it turns out that the contribution of high-z AGNs
to the ionizing background is probably not negligible at $z\sim 5$.

\subsection{The QSO Luminosity Function at $z\sim 6$}

The redshift evolution of the bright side of the QSO luminosity
function from $z\sim 3$ to $z\sim 5$ is mild, with a best-fit
parameter of $\gamma=-0.25$, as shown in Fig. \ref{fig:rhoz}. The
redshift evolution of the faint side from $z\sim 3$ to $z\sim 5$ is
similar to the bright one, as can be seen from this figure. At
$M_{1450}=-28.5$ there seems to be an accelerated evolution of the QSO
space density from $z\sim 5$ to $z\sim 6$, with respect to the mild
decrease observed at $z<5.5$. This trend is evident with the drop of
the bright end of the QSO luminosity function derived by
\citet{Jiang2016}. It is not clear, however, whether such a drop is
due to a physical reason (faster accretion of the SMBH population) or,
alternatively, if it is the effect of a strong incompleteness of past
surveys. In the future it will be important to extend the present QSO
surveys \citep{Sch19a,Sch19b,Wolf20,boutsia21,Onken21} at $z\sim 6$ in
order to constrain the redshift evolution of the QSO luminosity
function in a larger redshift interval.

It is interesting to note here that \citet{calverley11} inferred a
mild decline in the emissivity of ionizing photons by roughly a factor
of 2 between $z=5$ and $z=6$, by combining the measurements of the
evolution of the MFP of ionizing photons with the evolution
of the photo-ionization rate. If the ionizing UVB is entirely produced
by QSOs and AGNs, then the result of \citet{calverley11} is fully
consistent with a progression of a mild evolution of QSO luminosity
function with $\gamma=-0.25$ even at $5<z<6$.

\subsection{Strong lensing magnification ?}

An alternative explanation for the high space density of QSOs at
ultra-bright magnitudes could be interpreted also by the effect of
strong lensing magnification of intrinsically fainter objects, as
proposed by \citet{PacucciLoeb20}. In the unlikely case that all the
14 QSOs of our sample are strongly magnified by large amplification
factors, then the photo-ionization rate computed above could be no
longer valid, and it is possible that the QSO population will be thus
unable to provide the majority of the HI ionizing photons to maintain
the IGM ionized at $z\sim 5$. Such conclusions are depending on the
exact amount and distribution of the magnification factors.

We have queried the GAIA EDR3 database \citep{gaia16,gaiaedr3} to
check whether any of the 14 QSOs in Table \ref{tab:qsoz5} has a GAIA
EDR3 counterpart within 3 arcsec, which could be an indication of
strong lensing magnification. It turns out that none of the 14 QSOs at
$z\sim 5$ has a nearby object. This could suggest that these QSOs are
not strongly lensed by foreground objects, but a more careful analysis
is needed before excluding strong lensing magnification as the reason
for their exceptional luminosities. The strong lensing hypothesis
could be verified in the future with high-resolution imaging of these
ultra-bright QSOs with JWST or with ELT.


\section{Conclusions}

The QUBRICS survey \citep{calderone19,Bou20,boutsia21,guarneri21}
turns out to be particularly efficient and complete in the selection of
ultra-bright QSOs at high redshift ($z>2.5$). Thanks to the extensive
spectroscopic confirmations carried out progressing with this survey,
and complementing our database with the results of other groups
\citep{Wolf20,Onken21}, a sample of 14 ultra-bright QSOs with
$M_{1450}\le -28.3$ at $4.5<z_{spec}<5.0$ has been assembled in Table
\ref{tab:qsoz5}. Out of these 14 QSOs, 5 objects have a photometric
redshift derived by the CCA technique of $z_{CCA}\ge 4.5$.
With these 14 QSOs,
the bright side of the luminosity function at $z\sim 5$ has been
derived, as shown in Fig. \ref{fig:lfz5} and in Table \ref{tab:lfobs}.

From the QSO Luminosity function at $z\sim 5$ in the QUBRICS footprint,
and comparing them with similar results at $3\lesssim z\lesssim 6$,
we can derive a number of conclusions, which we summarize here:
\begin{enumerate}
\item
At z=4.75 the QSO space density in the absolute magnitude range
$-29.3\le M_{1450}\le -28.3$ is
$\Phi=3.115^{+1.077}_{-0.825}\times 10^{-10}cMpc^{-3}$.
\item
This space density of QSOs at $z\sim 5$ and $M_{1450}=-28.6$ is a factor of 3
higher than previous determination in the past, as also recently derived by
\citet{Sch19a,Sch19b,boutsia21} at $z\sim 3-4$ and by \citet{Onken21}
at $z\ge 4.5$.
This provides strong evidences that the previous results, mainly derived by the 
SDSS survey, may suffer by significant incompleteness, or their completeness
corrections have been underestimated.
\item
The bright-end slope of the $z\sim 5$ QSO luminosity function
is relatively steep, with a
best fit value of $\beta=-4.64$, and an upper limit
of $\beta\le -3.4$ at 68\% confidence level. There is no indication of a
flattening of the bright-end slope of the QSO luminosity function
going at high redshifts, as suggested by previous surveys.
\item
The observed space density of ultra-bright QSOs at $z\sim 5$
is not compatible with the recent
best-fit parameterizations shown in Fig. \ref{fig:lfz5},
which are underestimated by a factor of $\sim 3$ at $M_{1450}\sim -28.5$.
\item
The redshift evolution of the space density of ultra-bright QSOs
($M_{1450}\sim -28.5$)
between z=4 and z=5 is milder than previous determinations.
A fit to the space densities at $M_{1450}\sim -28.5$ in the redshift interval
$3\le z\le 5$ gives a mild logarithmic slope of $\gamma=-0.25$
(see Fig. \ref{fig:rhoz}).
\item
The evolution of the AGN space densities at much fainter luminosities of
$M_{1450}\sim -23$ yields a similar best-fit parameter of $\gamma=-0.27$,
indicating that at $z>3$ the evolution of the QSO luminosity function
is consistent with a pure density evolution law, i.e. a rigid shift
towards lower values at higher redshift, with a uniform mild decline
in the number density, independent of QSO luminosity.
\item
Adopting a pure density evolution, we have evolved the QSO luminosity
function of \citet{boutsia21} from $z=3.9$ to $z=4.75$, assuming
$\gamma=-0.25$.
The resulting luminosity function agrees with the QSO space density from
QUBRICS, as shown by the blue continuous curve in Fig. \ref{fig:lfz5}.
\item
The photo-ionization rate produced by bright QSOs and faint AGNs
at $z\sim 5$ in the absolute magnitude range $-30\le M_{1450}\le -18$ is
$\Gamma_{HI}=0.46^{+0.17}_{-0.09}\times 10^{-12} s^{-1}$, assuming an escape
fraction of 70\% and a steep faint-end slope of the luminosity function
($\alpha=-1.85$) at $z\sim 4$, as derived by \citet{glikman11} and \citet{boutsia21}.
This value of $\Gamma_{HI}$ produced by AGNs is $\sim$50-100\% times the 
ionizing UV background measured at $z\sim 5$ through different methods
(e.g. Lyman-$\alpha$ forest, Proximity effect, QSO near-zone size), as shown
in Table \ref{tab:gammaz5}.
\end{enumerate}

This indicates that QSOs at $z\sim 5$ do not have a marginal role in the
production of ionizing photons detected at such high redshift, provided that
their LyC escape fraction is high ($\ge 70\%$) and the faint end
of the luminosity function is rather steep.
Based on these results, a clear revision of the role of the AGN population
in the cosmological reionization of hydrogen should be carried out.
Paradoxically, the most enigmatic dark objects in the Universe, i.e. the
SMBHs, could significantly contribute to the so-called First Light
or Cosmic Dawn, ending the so called Dark Ages.

Future spectroscopic wide-field instrumentation (e.g. WEAVE, 4MOST,
Euclid) coupled with state-of-the art imaging surveys
(e.g. Rubin-LSST, Euclid, Roman Space Telescope) will extend the
present analysis to lower luminosities and higher redshifts, providing
a sound statistical sample of high-z QSOs. Armed with these data-sets,
it will be possible to study in detail the escape fraction of faint,
high-z AGNs and the luminosity function of QSOs
up to the Epoch of Reionization, and possibly even beyond.


\begin{acknowledgments}
We thank the referee for the useful comments that allowed us to improve the
quality of the paper.

AG and FF acknowledge support from PRIN MIUR project ‘Black Hole winds
and the Baryon Life Cycle of Galaxies: the stone-guest at the galaxy
evolution supper’, contract 2017-PH3WAT.

Part of the results discussed in this work are based on observations
made with the Italian Telescopio Nazionale Galileo (TNG) operated on
the island of La Palma by the Fundacion Galileo Galilei of the INAF
(Istituto Nazionale di Astrofisica) at the Spanish Observatorio del
Roque de los Muchachos of the Instituto de Astrofisica de Canarias
during periods AOT42 and AOT43.

This work is based on data products from observations made
with ESO Telescopes at La Silla Paranal Observatory under
ESO programmes ID 103.A-0746(A), 0103.A-0746(B), and
0104.A-0754(A).

This work has made use of data from the European Space Agency (ESA)
mission {\it Gaia} (\url{https://www.cosmos.esa.int/gaia}), processed
by the {\it Gaia} Data Processing and Analysis Consortium (DPAC,
\url{https://www.cosmos.esa.int/web/gaia/dpac/consortium}).  Funding
for the DPAC has been provided by national institutions, in particular
the institutions participating in the {\it Gaia} Multilateral Agreement.

This paper includes data gathered with the 6.5 meter Magellan
Telescopes located at Las Campanas Observatory (LCO), Chile.

The national facility capability for SkyMapper has been funded through
ARC LIEF grant LE130100104 from the Australian Research Council,
awarded to the University of Sydney, the Australian National
University, Swinburne University of Technology, the University of
Queensland, the University of Western Australia, the University of
Melbourne, Curtin University of Technology, Monash University and the
Australian Astronomical Observatory. SkyMapper is owned and operated
by The Australian National University's Research School of Astronomy
and Astrophysics. The survey data were processed and provided by the
SkyMapper Team at ANU. The SkyMapper node of the All-Sky Virtual
Observatory (ASVO) is hosted at the National Computational
Infrastructure (NCI). Development and support the SkyMapper node of
the ASVO has been funded in part by Astronomy Australia Limited (AAL)
and the Australian Government through the Commonwealth's Education
Investment Fund (EIF) and National Collaborative Research
Infrastructure Strategy (NCRIS), particularly the National eResearch
Collaboration Tools and Resources (NeCTAR) and the Australian National
Data Service Projects (ANDS).

This publication makes use of data products from the Wide-field
Infrared Survey Explorer, which is a joint project of the University
of California, Los Angeles, and the Jet Propulsion
Laboratory/California Institute of Technology, funded by the National
Aeronautics and Space Administration.

This publication makes use of data products from the Two Micron All
Sky Survey, which is a joint project of the University of
Massachusetts and the Infrared Processing and Analysis
Center/California Institute of Technology, funded by the National
Aeronautics and Space Administration and the National Science Foundation.
\end{acknowledgments}

\facilities{SkyMapper, Wise, Gaia, Magellan:Baade (IMACS),
Magellan:Clay (LDSS-3), TNG:Dolores, NTT:Efosc2}


\bibliography{ReferencesLF}{}
\bibliographystyle{aasjournal}

\end{document}